# New tool for extraction of $^{187}$Os Mössbauer parameters with biologically relevant detection sensitivity

Iryna Stepanenko,[1,2] Zhishuo Huang,[3] Liviu Ungur,[3] Dimitrios Bessas,[4] Aleksandr Chumakov,[4] Ilya Sergueev,[5] Gabriel E. Büchel,[6] Abdullah A. Al-Kahtani,[7] Liviu F. Chibotaru,[8,*] Joshua Telser,[9,*] Vladimir B. Arion[1,*]

[1]*University of Vienna, Institute of Inorganic Chemistry, Währinger Strasse 42, 1090 Vienna, Austria*

[2]*Institute of Science and Technology Austria (ISTA), Am Campus 1, A-3400 Klosterneuburg, Austria*

[3]*Department of Chemistry, National University of Singapore, Block S8 Level 3, 3 Science Drive 3, 117543, Singapore*

[4]*European Synchrotron Radiation Facility, F-38043 Grenoble, France*

[5]*Deutsches Elektronen Synchrotron D-22607 Hamburg, Germany*

[6]*ChemConsult GmbH, P.O. Box 43, 9485 Nendeln, Liechtenstein*

[7]*Chemistry Department, College of Science, King Saud University, P.O. Box 2455, Riyadh 11451, Saudi Arabia*

[8]*Theory of Nanomaterials Group, KU Leuven, Celestijnenlaan 200F, B-3001 Leuven, Belgium*

[9]*Department of Biological, Physical and Health Sciences, Roosevelt University, 430 S. Michigan Avenue, Chicago, Illinois 60605, USA*

**Abstract**. A large number of osmium complexes with osmium in different oxidation states (II, III, IV, VI) have been reported recently to exhibit good antiproliferative activity in cancer cell lines. Herein, we demonstrate new opportunities offered by $^{187}$Os nuclear forward scattering (NFS) and nuclear inelastic scattering (NIS) of synchrotron radiation for characterization of hyperfine interactions and lattice dynamics in a benchmark Os(VI) complex K$_2$[OsO$_2$(OH)$_4$], by accurate extraction of Mössbauer parameters and the determination of Os-projected density of phonon states confirmed by first-principles phonon calculations. The values of isomer shift ($\delta$ = 3.3(1) mm/s) relative to [Os$^{IV}$Cl$_6$]$^{2-}$ and quadrupole splitting ($\Delta E_Q$ = 12.0(2) mm/s) were determined with NFS, while the Lamb-Mössbauer factor (0.55(1)), the density of phonon states (DOS), and a full thermodynamics characterization was carried out using the NIS data combined with first principle theoretical calculations. In more general terms, this study provides strong evidence that $^{187}$Os nuclear resonance scattering is a reliable technique for the investigation of hyperfine interactions and Os specific vibrations in osmium(VI) species, which might be potentially applicable for measuring such interactions in osmium complexes of other oxidation states, including those with anticancer activity such as Os(III) and Os(IV).

**Introduction**

The use of metal complexes as anticancer agents revolutionized cancer treatment more than fifty years ago with the use of Cisplatin, *cis*-[PtCl$_2$(NH$_3$)$_2$].[1] The efficiency of Pt complexes is in part due to their specific kinetics towards ligands substitution and rearrangement, which allows kinetically controlled metal binding to DNA.[2] Other metal ions of the so-called platinum group (Ru, Rh, Os, Ir, Au) that have similar ligand-exchange kinetics, have also attracted the attention of researchers in a hope to further increase the efficacy and reduce the number of side effects and general toxicity of Pt-based drugs. Nowadays Ru and Os compounds are being considered as analogous anti-cancer drugs, with studies on osmium complexes being more recent. Two ruthenium(III) complexes, namely KP1019, indazolium *trans*-[tetrachloridobis(1*H*-indazole)ruthenate(III)], and NAMI-A, imidazolium *trans*-[tetrachlorido(dimethyl sulfoxide)(1*H*-imidazole)] have entered clinical trials for treatment of a broad range of anti-cancer indications and metastases, respectively.[3,4] This potential for Ru-based drugs has also fueled interest in analogous osmium complexes. The advantage of using osmium analogues with a cytotoxicity similar to that of their ruthenium congener lies in the higher inertness of osmium species under conditions relevant for drug formulation.[5,6] A family of azole complexes with osmium in different oxidation states (II, III, IV, VI)[5,7,8,9,10,11,12] and osmium(VI) complexes with Schiff-base ligands[13] have been reported recently, in addition to a large number of organoosmium(II) arene compounds,[14,15,16,17,18,19,20,21,22] which exhibited varied antiproliferative activity.

The mechanisms of action of many Ru- and Os-based anticancer drug candidates remain unclear. Among other reasons, this is due to a lack of metallodrug speciation data in physiologically relevant conditions,[23,24,25] including aqueous buffers, protein solutions, undiluted blood serum, cell culture medium and human cancer cells. The complexity of biological fluids makes speciation studies using classical spectroscopic methods (Nuclear Magnetic Resonance (NMR), Electron Paramagnetic Resonance (EPR), or Ultraviolet–visible Spectroscopy) or ElectroSpray Ionisation-Mass Spectrometry (ESI-MS) challenging. Notable efforts have been made by Walsby and co-workers to apply EPR spectroscopy to the study of clinically relevant Ru(III) complexes under physiological conditions.[26,27,28,29,30,31] Atomic Absorption Spectroscopy (AAS) and Induced Coupled Plasma-Mass Spectrometry (ICP-MS) are destructive and provide only total metal amount without information on metal speciation unless individual components are separated.[24] Lay and co-workers demonstrated that X-ray absorption spectroscopy (XAS) is superior for having elemental specificity, being nondestructive to drugs, and having greater tolerance for biologically relevant matrices.[24] All these advantages make XAS the currently preferred technique for the speciation of metallodrugs in biological systems. By using XAS, strong evidence was obtained that the antimetastatic activity of NAMI-A is due to formation of adducts with serum albumin, in which the original complex did not preserve its integrity and lost its imidazole ligands.[32] KP1019 was shown to react with proteins with formation of $Ru^{IV/III}$ clusters and binding of $Ru^{III}$ to S donor atoms, amine-imine, and carboxylate groups.[25] The original KP1019 complex was suggested to be taken into the cells by passive diffusion and not by an active cellular uptake binding pathway via transferrin as was suggested previously.[23] Even though the internalization of KP1019 by cancer cells was found to be 20 times higher than that of NAMI-A, corelating with the low cytotoxicity of the latter, the cell uptake of KP1019 reduced markedly after incubation in cell-culture media due to the complex's decomposition.[23]

One disadvantage of XAS[23,24,25] (and EPR[26,27]) is that often higher concentrations than those biologically relevant should be used in the cell medium to obtain a detectable cellular response. Therefore, extrapolation of any such results to biologically relevant concentrations requires additional care as the speciation may change. In this context the development of new element-specific techniques with detection sensitivity at the level of biologically relevant concentrations is desirable.

Herein, we report a benchmark osmium-specific characterization of potassium osmate, $K_2[OsO_2(OH)_4]$, by $^{187}Os$ nuclear resonance scattering of synchrotron radiation using very small amounts of $^{187}Os$ (68 µg/mm$^2$) compared to 51 mg Ru typically used in Ru K-edge XAS studies.[25]

The isomer shift ($\delta$) and the nuclear quadrupole splitting ($\Delta E_Q$) are extracted and the Os specific density of vibrational states (DOS) is determined and compared with that obtained by first-principles theoretical calculations, which allowed for a precise assignment of vibrational modes containing significant displacements of the Os atom. These results show that nuclear resonance scattering of synchrotron radiation applied to $^{187}$Os species is a powerful tool for measuring local electronic and vibrational properties in osmium complexes, thus offering unprecedented opportunities for speciation studies of osmium-based anticancer drugs.

**Experimental**

**Synthesis.** K$_2$[$^{187}$OsO$_2$(OH)$_4$] was prepared from $^{187}$Os metal (99.55% enriched, obtained from SC "PA Electrochemical Plant", Russia, which was converted to $^{187}$OsO$_4$ by heating in a quartz tube in a stream of air. Enriched osmium tetroxide was reacted with 6 M aqueous KOH followed by addition of ethanol as reductant. Crystallization in air afforded X-ray diffraction quality single crystals. The determined cell parameters at 100 K for a tetragonal crystal, *I*4/*mmm* (*a* = *b* = 5.5904(2), *c* = 9.4276(4) Å) were in line with those reported in the literature.[33] The three strong IR absorption bands at 3281, 1105, and 795 cm$^{-1}$ (Figure S1) are in good agreement with those reported for the same complex with naturally abundant osmium[33] providing further evidence for sample purity. The negative ion ESI mass spectrum provides further evidence of its identity. A peak with *m/z* 328.88 could be assigned to the ionic pair {K$^+$[$^{187}$OsO$_2$(OH)$_4$]$^{2-}$}$^-$ (Figure S2). (H$_2$pz)$_2$[$^{187}$OsCl$_6$] was prepared as an orange powder by reaction of [(DMSO)$_2$H]$_2$[$^{187}$OsCl$_6$], which is prepared directly from $^{187}$OsO$_4$,[9] in dry ethanol with an excess of 1*H*-pyrazole at room temperature as reported previously.[11] The negative ion ESI mass spectrum showed a peak with *m/z* 363.75, which could be attributed to [$^{187}$Os$^{IV}$Cl$_5$]$^-$ (see Figure S3).

**Sample preparation**. Polycrystalline K$_2$[$^{187}$OsO$_2$(OH)$_4$] (~8 mg) was placed in an area of 10 mm$^2$ and enclosed in Kapton® tape.

**Hyperfine interaction and lattice dynamics characterization**. Nuclear resonance scattering measurements, both Nuclear Forward Scattering (NFS) and Nuclear Inelastic Scattering (NIS), were carried out at Nuclear Resonance beamlines ID18[34] and ID14 of the European Synchrotron Radiation Facility (ESRF), Grenoble, France. The synchrotron radiation storage ring was operating in 16-bunch mode providing X-ray flashes every 176 ns. The optical elements used at the nuclear resonance energy of $^{187}$Os, 9.778(3) keV, have been described recently.[35] For acquiring both the NFS and NIS spectra sets of avalanche photo diodes were used. Both NFS and NIS measurements were carried out at ambient conditions. The samples (see above) were placed on motorized stages and a region with a thickness that corresponds to approximately one electronic absorption length was identified and chosen for the NFS measurements.

**Computational methods.** All electronic and phonon structure calculations were performed by Quantum ESPRESSO[36,37] with the revised Perdew-Burke-Ernzerhof (PBE) functional[38] and optimized norm-conserving Vanderbilt pseudopotentials[39] taken from the pseudopotentials PSEUDO DOJO.[40] The plane-wave kinetic energy cutoff and the density cutoff were set to 1360 eV and 5442 eV, respectively, and a shifted 8 × 8 × 8 Monkhorst–Pack mesh was employed for Brillouin zone integration in order to ensure convergence. The convergence threshold of total energy was $1.3 \times 10^{-11}$ eV for SCF calculations. The experimental structure was fully relaxed with the force on each atom smaller than convergence and the total energy convergence threshold set to $2.6 \times 10^{-4}$ eV/Å and $1.3 \times 10^{-13}$ eV, respectively. Phonon calculations (Figures S4 and S5) were performed with density functional perturbation theory (DFPT),[41] in which a 2 × 2 × 2 mesh for Brillouin zone integration and $1.3 \times 10^{-15}$ eV for the self-consistent threshold were used. The projected phonon DOS calculation was carried out with a 28 × 28 × 28 mesh to guarantee convergence. Due to the existence of Os–OH bonding in the ion [OsO$_2$(OH)$_4$]$^{2-}$, the material can be considered as a pseudo molecular crystal, thus the non-local interaction (vdW) should be taken into consideration, which is performed by the vdW-DF method.[42,43]

**Results**

The K$_2$[$^{187}$OsO$_2$(OH)$_4$] was prepared as described in the literature and its purity was confirmed by comparison of cell parameters of the single crystals and Attenuated Total Reflectance (ATR) IR spectrum (see Figure S1). The X-ray diffraction structure of K$_2$[OsO$_2$(OH)$_4$] is shown in Figure 1.[33]

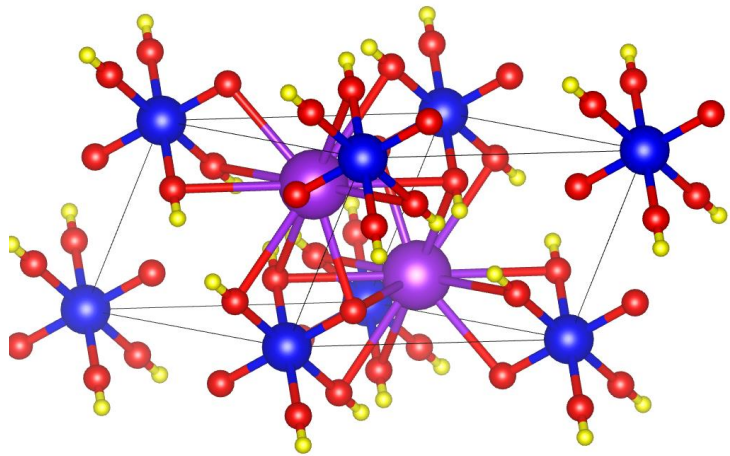

**Figure 1.** Crystal structure of $K_2[OsO_2(OH)_4]$. Blue, purple, red, and yellow spheres represent Os, K, O, and H, respectively.

*Hyperfine interactions.* Time evolution of nuclear decay for $K_2[^{187}OsO_2(OH)_4]$ was measured by NFS in the time interval between 6 and 41 ns after the prompt pulse of incident radiation. Typical count rate was about 1000 photons per second. Such a count rate allows one to measure a NFS spectrum with reasonable statistics in about an hour. The NFS data for $K_2[^{187}OsO_2(OH)_4]$ shown in Figure 2 depicts the usual exponential decay superimposed onto clear oscillations with a period of ~10 ns. The data were fitted between 13 and 40 ns using the program CONUSS[44,45] by a model including a quadrupole hyperfine interaction with the quadrupole splitting, $\Delta E_Q = 12.0(2)$ mm/s and the texture parameter responsible for the preferable orientation of the electric field gradient direction equal to 39±2%. The NFS data for times earlier than 13 ns were not included in the fit since they were heavily distorted due to overloading of the avalanche photodiode from the prompt X-ray flashes coming from the synchrotron storage ring.

For extracting the isomer shift (δ) a NFS measurement relative to a standard sample is required. In this case $(H_2pz)_2[^{187}Os^{IV}Cl_6]$ (Hpz = 1*H*-pyrazole) was chosen as the standard sample because its Os environment is fully isotropic, *i.e.*, no quadrupole splitting is expected for $(H_2pz)_2[^{187}Os^{IV}Cl_6]$. Three NFS measurements were carried out in this case: (i) a NFS measurement of $K_2[^{187}OsO_2(OH)_4]$ alone, (ii) a NFS measurement of $(H_2pz)_2[^{187}Os^{IV}Cl_6]$ alone (see Supporting Information, Figure S6), and (iii) a NFS measurement of $K_2[^{187}OsO_2(OH)_4]$ and $(H_2pz)_2[^{187}Os^{IV}Cl_6]$ simultaneously (see Supporting Information, Figure S7). Special attention was taken to measure NFS at the same points of the $K_2[^{187}OsO_2(OH)_4]$ and the $(H_2pz)_2[^{187}Os^{IV}Cl_6]$ samples during the individual and the combined NFS measurements. Indeed, the NFS measurement of $(H_2pz)_2[Os^{IV}Cl_6]$ shows only a simple exponential decay, without the presence of additional beat(s) which indicates that $\Delta E_Q = 0$ for $(H_2pz)_2[Os^{IV}Cl_6]$, as expected. After extracting all relevant information from the individual NFS measurements, the isomer shift ($\delta$) was the only unknown parameter in the combined measurement. The isomer shift ($\delta$) for $K_2[^{187}OsO_2(OH)_4]$ is found to be 3.3(1) mm/s relative to $(H_2pz)_2[^{187}Os^{IV}Cl_6]$.

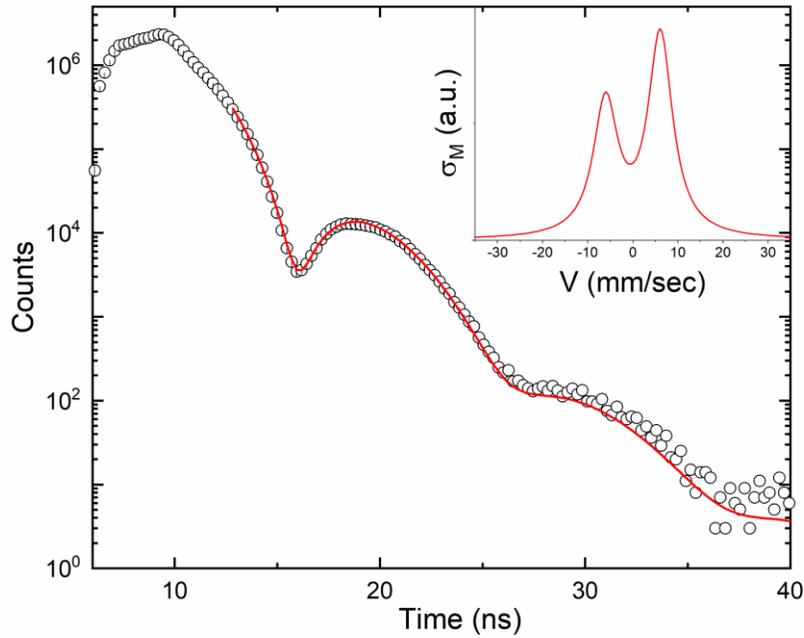

**Figure 2**. Time evolution of $^{187}Os$ nuclear decay measured with NFS for $K_2[^{187}OsO_2(OH)_4]$. The red curve shows the fit by CONUSS software. Standard deviation, σ, follows Poisson statistics, thus, a typical error bar (not given above) may be calculated as the square root of the recorded intensity. The inset shows simulation of the Mössbauer cross-section versus Doppler drive velocity corresponding to the fit.

*Lattice dynamics.* The NIS measurements of $K_2[^{187}OsO_2(OH)_4]$ were recorded between –10 and +60 meV with respect to the $^{187}Os$ transition energy (9.778(3) keV). The Os specific density of vibrational states in $K_2[^{187}OsO_2(OH)_4]$ is extracted by employing the double Fourier transformation as implemented in the software DOS.[46] The resulting normalized (to unit area) Os specific density of vibrational states in $K_2[^{187}OsO_2(OH)_4]$ is depicted in Figure 3. Two major peaks are observed between 5 and 20 meV followed by two minor peaks between 35 and 44 meV. The density of vibrational states, g(E), provides direct access to a series of thermodynamic parameters.[47] Firstly, the probability of the recoilless absorption, known as Lamb-Mössbauer factor, can be extracted using Equation (1),

$$f_{LM} = \exp\left(-E_R \int g(E)\left(\frac{1+e^{-\beta E}}{1-e^{-\beta E}}\right)\frac{dE}{E}\right), \quad (1)$$

where $E_R$ = 0.274 meV, is the recoil energy for an $^{187}$Os isolated nucleus and $\beta = 1/k_B T$, where $k_B$ is the Boltzmann constant and $T$ is the temperature at which g(E) is measured. The calculated $f_{LM}$ for Os in K$_2$[$^{187}$OsO$_2$(OH)$_4$] is 0.55(1) at room temperature. From the Lamb-Mössbauer factor, the purely incoherent mean-square atomic displacement parameter, $\langle u^2_{Os}\rangle$, is

$$\langle u^2_{Os}\rangle = \frac{-\ln(f_{LM})}{k^2}, \quad (2),$$

where $k$ = 4.959 Å$^{-1}$ is the wave number of the resonant photons. The Os mean square atomic displacement in K$_2$[$^{187}$OsO$_2$(OH)$_4$] at room temperature is 243(1) pm$^2$. The Os specific mean-force constant, $\langle F_{Os}\rangle$ is obtained from Equation (3),

$$\langle F_{Os}\rangle = M\int \frac{g(E)E^2 dE}{\hbar^2}, \quad (3),$$

where $M$ is the mass of the resonant isotope, 187 amu. The obtained value for K$_2$[$^{187}$OsO$_2$(OH)$_4$] is 387(1) N/m. In addition, the density of vibrational states gives the vibrational entropy, $S^{vib}_{Os}$ as obtained from Equation (4), using hyperbolic trigonometric functions.

$$S^{vib}_{Os} = 3k_B \int_0^\infty g(E)\left[\frac{\beta E}{2}\coth\left(\frac{\beta E}{2}\right) - \ln\left(2\sinh\left(\frac{\beta E}{2}\right)\right)\right] dE \quad (4).$$

The contribution of the Os specific vibrations to the total entropy of the system is 5.25(1) $k_B$.

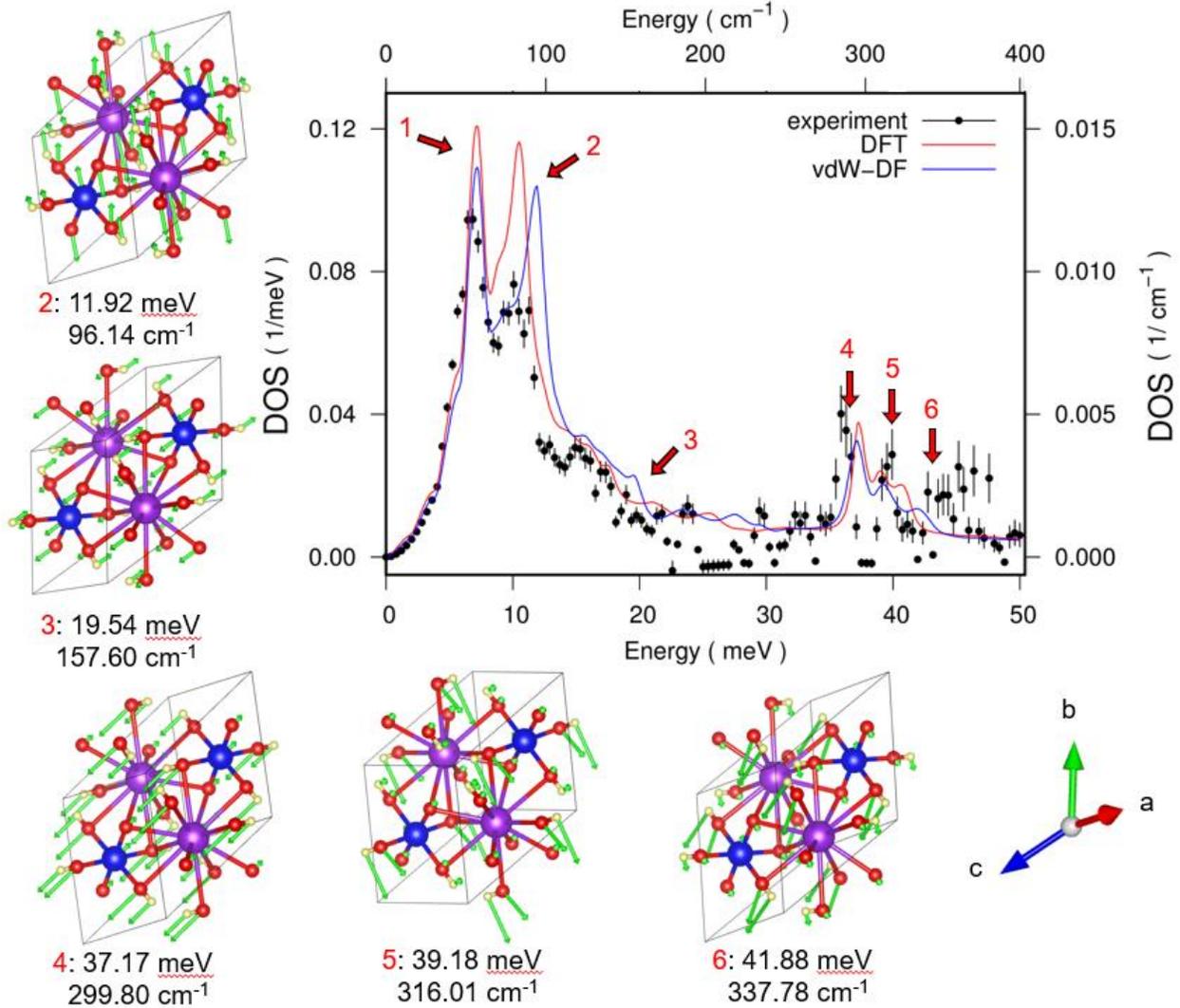

**Figure 3**. Os-projected phonon density of states (DOS) of $K_2[^{187}OsO_2(OH)_4]$ from NIS (black circles) and first-principles calculations (lines). Red and blue lines represent DFT calculations with a PBE functional and with additional vdW corrections, respectively. The structural diagrams surrounding the central figure show atomic displacements (green arrows) at each corresponding Os-projected DOS peak (red numbers in DOS plot and displacement diagrams) at the Γ point. Peak 1 corresponds to a single transverse acoustic mode and thus has no corresponding atomic displacement diagram. The atom coloring scheme is as in Figure 1: blue, purple, red, and yellow spheres represent Os, K, O, and H, respectively. The numbers under each plot give the position (energy) of the peak, corresponding to the numbered red arrows in the Os-projected phonon density of states.

The calculated Os-projected phonon DOS is shown in Figure 3 by blue and red lines corresponding to the two employed calculation methods. To facilitate comparison with experimental data (black circles in Figure 3), the calculated Os-projected phonon DOS (Figure S4) was convoluted with the instrumental function of the employed high-resolution monochromator. The corresponding atomic displacements at the Γ point, which is the highest symmetry point (the zero-momentum transfer point; also known as the center of the Brillouin zone), are shown in the lateral plots in Figure 3. We can see that there are in total six peaks, of which two peaks are in the lower-energy region (<12 meV) with much higher strength than the four peaks in the higher energy region (19–42 meV). This is because of the large atomic mass difference between Os and the other atoms. The first peak shown in Figure 3 corresponds to a single transverse acoustic (TA) mode. The other five peaks represent Os-involved optical modes with the representative vibrations shown as the surrounding subplot in Figure 3. Peak 2 corresponds to the vibration between the group of Os+(OH) and that of K+O and also shows contributions from longitudinal acoustic (LA) and second TA modes. Peak 3 corresponds to the displacement of the group of Os+O in (OH) relative to the group K+O+H in (OH). Peak 4 involves the vibration between the group of Os+K+O and OH. Peaks 5 and 6 correspond to vibrations involving the three groups: Os+K, O+(OH), and (OH).

Peaks 2, 3, and 6 calculated by DFT differ slightly from those calculated by the vdW+DF method. Such a difference is reasonable, as vdW+DF considers non-local interaction corrections. Peak 1 does not involve sensible relative displacements of nearby atoms, thus the two methods predict the same peak energy. For Peak 2, since the two groups involved in vibration, Os+(OH) and K+O, have a largely spaced distribution of charge, the non-local interaction makes a shift of about 2 meV. The situation with Peaks 3 and 6 is as with Peak 2, involving additional local vibration of O-H. Peak 4 corresponds to the displacements of individual atoms (Os, K, and O) and (OH) groups, all having localized charge distributions, therefore, DFT and vdW-DF calculations give a similar result, which is alike to the situation of Peak 5.

**Discussion**

In addition to XAS, synchrotron-based X-ray fluorescence (XRF) spectroscopy has been found recently to be a valuable tool for the investigation of biotransformations of anticancer ruthenium(III) and organoosmium(II) complexes.[25,48] Nuclear resonance techniques, such as Ru or Os NMR and Mössbauer spectroscopies, have been employed even more rarely, if at all.

Multinuclear NMR is a powerful technique for investigating metal-based drug speciation under physiologically relevant conditions in case of diamagnetic species (e.g., low-spin d$^6$ complexes).[49] Direct observation of NMR resonances for the metals themselves is potentially very useful, but often difficult to achieve. Neither of the two magnetically active isotopes of Os, $^{187}$Os ($I$ = 1/2) and $^{189}$Os ($I$ = 3/2) with 1.64% and 16.10% natural abundancies, respectively, (see https://nmr.wsu.edu/nmr-periodic-table/) provide a reasonable NMR signal. In the case of $^{187}$Os, not only is the abundance low, but the nuclear magnetic moment is very small (resonance at only 11.5 MHz for 600 MHz NMR spectrometer). For $^{189}$Os (https://www.webelements.com/osmium/isotopes.html) the difficulty is the quadrupolar nucleus with a large quadrupole moment leading to broad NMR lines unless in highly symmetric molecules, e.g., OsO$_4$. At the same time, indirect detection of $^{187}$Os nuclear resonances by polarization transfer techniques from sensitive nuclei has been reported.[50] This allows for a great increase in sensitivity of $^{187}$Os detection. Sharp NMR resonances were reported for a large number of osmium-arene complexes with phosphines and cyclopentadienyl (Cp) complexes,[51,52] μ-hydrido,[53] [Os(η$^6$-biphenyl)$_2$](CF$_3$SO$_3$)$_2$,[54] by using polarization transfer techniques via $^{187}$Os–$^1$H and $^{187}$Os–$^{31}$P couplings. The isotopic enrichment of [Os((η$^6$-$p$-cymene)($N$,$N$-azpy-NMe$_2$)Br]PF$_6$ to more than 98% allowed straightforward detection of its $^{187}$Os resonance by using two-dimensional $^{187}$Os–$^1$H HMBC and $^1J$, $^2J$, $^3J$, and $^4J$ coupling constants for both the arene ring and the coordinated azopyridine ligand.[49]

In contrast to NMR or EPR, Mössbauer spectroscopy can be applied to species with any number of d electrons, even, odd, or none. This technique is highly effective in determining the electronic structure and oxidation state of iron complexes, using $^{57}$Fe.[55] Further within Group 8, examples of $^{99}$Ru ($I$ = 5/2, 12.76% abundance) Mössbauer investigations are well-documented,[56,57,58,59,55] but radiochemical (cyclotron) facilities for the $^{99}$Rh precursor (half-life of 16 days) are currently unavailable and the high energy (89.36 keV) is a complicating issue in the use of synchrotron radiation.[60] In particular, the high energy of the $^{99}$Ru Mössbauer transition leads to a low Lamb-Mössbauer factor which thus requires low temperature NFS measurements, while NIS measurements at such a high energy cannot achieve meV resolution. The same Os isotopes as for NMR spectroscopy are of interest for Mössbauer spectroscopy. $^{189}$Os Mössbauer spectra were successfully measured for a series of 15 osmium compounds in oxidation states +8, +6, +4, +3, and +2 by using the best suited 36.2 and the 69.5 keV nuclear transitions.[61] However, at that early date no Os(II), Os(III), Os(IV), and Os(VI) coordination complexes of relevance as anti-cancer agents were studied; indeed, this work was done before the introduction even of Pt-based anticancer drugs. The lifetime of the 36.2 keV level ($\tau$ = 0.74 ns)[62,63] is markedly shorter than that of the 69.5 keV level ($\tau$ = 2.3 ns).[64] The corresponding minimum FWHM (full width half maximum) Mössbauer linewidths, $W_0 = 2h/\tau$, are 15 mm/s for the 36 keV transition and 2.5 mm/s for the 69.5 keV transition. Nevertheless, the 36.2 keV transition is more convenient for isomer shift studies than the 69.5 keV transition, since the change of the mean-square nuclear charge radius, $\Delta\langle r^2\rangle$, for the 36.2 keV transition exceeds that for the 69 keV transition by a factor of 16.[65] On the other hand, the 36.2 keV Mössbauer line is too broad for any but the largest electric quadrupole and magnetic dipole hyperfine splittings to be even partially resolved. In any case, the short half-life of the radioactive source, $^{189}$Ir, is only 13.3 days. This makes such studies for a broader research community inconvenient from practical point of view and no such sources are currently available. The use of $^{187}$Os isotope for Mössbauer spectroscopy would also be of great interest, but there is no suitable radioactive source.

This unfortunate situation was recently lifted by successful excitation[35] of the low lying first excited nuclear level energy state for $^{187}$Os at 9.778(3) keV by synchrotron radiation, with demonstration of NFS and NIS feasibility. This has now opened the way to lattice dynamics and hyperfine characterization of osmium complexes by nuclear forward scattering (NFS)[66,67] and nuclear inelastic scattering (NIS),[68,69] namely experimental determination of the density of phonon states and extraction of Mössbauer parameters. High quality NIS spectra for metallic osmium allowed for experimental determination of DOS. The combination of low energy of nuclear transition and the large nuclear mass of $^{187}$Os resulted in a high recoil free fraction, $f_{LM}$ = 0.95(1), at room temperature which makes nuclear resonance scattering highly attractive for investigation of electronic and elastic properties of osmium complexes. However, no application of this technique for extraction of Mössbauer parameters and DOS has yet been reported for osmium complexes.

Herein we apply this new tool for investigation of $K_2[^{187}OsO_2(OH)_4]$ (Figure 1) in which Mössbauer parameters and DOS are indeed determined. The osmium(VI) in the complex is coordinated to four hydroxido oxygen atoms in the equatorial plane at 1.99(2) Å and two oxido oxygen atoms at shorter interatomic distance (1.75(2) Å) due to strong Os–O π-interactions.[33] Since Os is the heaviest atom in $K_2[^{187}OsO_2(OH)_4]$, according to the mass homology relation[70] it contributes to the lower energy phonon branches, i.e., energies not exceeding 50 meV (see Figure S5), albeit the entire phonon spectrum extends up to energies > 400 meV (see Figure S4).

The central plot in Figure 3 shows several prominent peaks in the calculated Os-pDOS. The first one (at ~7 meV) is of acoustic origin while the others originate from optical vibrations (Figure 3, outer diagrams). We note that the calculated phonon DOS at 0 K matches well the experimental one measured at room temperature. The comparison of the two calculational approaches (Figure 3) shows that the vdW-DF method gives a better agreement with experiment.

This study is a benchmark for future investigations. For example, the $K_2[^{187}OsO_2(OH)_4]$ was converted to other two osmium(VI) complexes, namely osmium(VI)-nitrido complexes with four equatorial chlorido ligands, which are medicinally more relevant, instead of four hydroxide ligands. Instead of axial oxido, these have either only one apical nitrido ligand or two axial ligands: nitrido and aquo. NFS and NIS studies of these will allow one to follow how the coordination of the sixth aquo ligand will affect the Os–N vibration, and how the Os–Cl vibrations will appear after replacement of hydroxido ligands.

One should note that besides Γ, the highest symmetry point, other symmetry points of the Brillouin zone contribute to the peaks due to the extremal nature of phonon dispersion in these lower symmetry points. Compared to calculations of vibrations in isolated molecules,[71] phonon calculations give a more realistic picture of atomic displacements at the Γ point because they take into account the interactions among molecular units in the crystal, which is especially important for low-frequency optical modes.[72]

It is well-documented, that the isomer shifts ($\delta$) for Fe, Ru, Os, Ir, Pt, and Au compounds are dependent on the oxidation state of the transition metal. Generally, an increase of $\delta$ correlates with an increase of s-electron density at the nucleus which in turn is related to an increase in oxidation state of the transition metal, which diminishes the deshielding effect of the remaining valence d-electrons. This is a general pattern, which, however, can be affected if ligands with back-bonding abilities, such as CO, $CN^-$, or $NO^+$, are involved in coordination to the metal. Transfer of electron density from d-orbitals to empty π* orbitals of the ligand can result in an increase of s-electron density on the nucleus, similar to what is seen with a higher formal oxidation state. The $\delta$ is also useful in determination of electron shielding and the electron-withdrawing power of electronegative groups on the ligands. Here, we found $\delta$ for K$_2$[$^{187}$OsO$_2$(OH)$_4$] to be 3.3(1) mm/s vs H$_2$pz[$^{187}$Os$^{IV}$Cl$_6$], which was used as a convenient standard for our systems of interest. We plan to expand the range of Os oxidation states in future work, allowing correlation of $^{187}$Os isomer shift with coordination chemistry.

The second parameter is the quadrupole splitting, which in addition to determination of the oxidation state, can be useful for identification of spin state, site symmetry, and the arrangement of ligands. Quadrupole splitting ($\Delta E_Q$) arises due to the interaction of the energy levels of the nuclei with a non-spherical charge distribution ($I > ½$) with an external electric field gradient, whose sign and strength depend on the ligand environment of the nuclei. In this case an asymmetrical electric field (produced by an asymmetric electronic charge distribution or ligand arrangement) splits the nuclear energy levels. The quadrupole splitting of K$_2$[$^{187}$OsO$_2$(OH)$_4$], $\Delta E_Q$ = 12.0(2) mm/s, clearly indicates that the Os local environment is anisotropic, as expected for this strongly axial system. These two Mössbauer parameters, which we show can now be easily obtained for $^{187}$Os, can often be used for identification of a particular species by comparison to spectra for model compounds. In addition, the relative intensities of the various peaks reflect the relative concentrations of compounds in a sample and can be used for semi-quantitative analysis.

**Conclusions**

Nuclear forward and nuclear inelastic scattering of synchrotron radiation by the low-lying nuclear level of $^{187}$Os (9.778(3) keV) in K$_2$[$^{187}$OsO$_2$(OH)$_4$] was investigated. The isomeric shift, $\delta$, and quadrupole splitting, $\Delta E_Q$, were determined by fitting the experimental NFS spectra. These results provide strong evidence that $^{187}$Os NFS is a reliable technique for investigation of hyperfine interactions in an osmium(VI) compound, and this approach can therefore be extended to complexes of Os(VIII), Os(IV), Os(III), and Os(II) as well as to other Os(VI) complexes of biological relevance. NIS spectroscopy performed with a 1 meV resolution allowed extraction of the density of phonon states, Os-pDOS, which in combination with first-principles phonon calculations, allowed the identification of vibrational modes containing significant displacements of Os. Expanding the library of the compounds investigated with NFS and NIS for benchmarking of the $\delta$ and $\Delta E_Q$ parameters for a series of osmium compounds with osmium in a series of oxidation states and varied octahedral ligand environment, will allow for performing speciation studies in frozen biological fluids of osmium-based drugs and provide insight into their mechanisms of anticancer activities both *in vitro* and *in vivo*.

## ASSOCIATED CONTENT

**Supporting Information**

The Supporting Information is available free of charge at https://pubs.acs.org/doi/10.1021/jacs

IR spectrum, ESI mass spectra, phonon dispersion and phonon DOS, phonon dispersion in low-energy domain and the corresponding Os-projected phonon DOS convoluted with the instrumental resolution density of states for $K_2[^{187}OsO_2(OH)_4]$.

## AUTHOR INFORMATION

**Corresponding Authors**

**Liviu Chibotaru** – *Theory of Nanomaterials Group, KU Leuven, Celestijnenlaan 200F, B-3001 Leuven, Belgium;* orcid.org/0000-0003-1556-0812; email: Liviu.Chibotaru@kuleuven.be

**Joshua Telser** – *Department of Biological, Physical and Health Sciences, Roosevelt University, 430 S. Michigan Avenue, Chicago, Illinois 60605, USA*; orcid.org/0000-0003-3307-2556; email: jtelser@roosevelt.edu

**Vladimir B. Arion** – *University of Vienna, Institute of Inorganic Chemistry, Währinger Strasse 42, A-1090 Vienna, Austria;* orcid.org/0000-0002-1895-6460; email: vladimir.arion@univie.ac.at

**Authors**

**Iryna Stepanenko** − *University of Vienna, Institute of Inorganic Chemistry, Währinger Strasse 38, A-1090 Vienna, Austria; Institute of Science and Technology Austria (ISTA), Am Campus 1, A-3400 Klosterneuburg;* https://orcid.org/0000-0002-9406-5539.

**Zhishuo Huang** − *Department of Chemistry, National University of Singapore, Block S8 Level 3, 3 Science Drive 3, 117543, Singapore;* orcid.org/0000-0002-9924-8465

**Liviu Ungur** – *Department of Chemistry, National University of Singapore, Block S8 Level 3, 3 Science Drive 3, 117543, Singapore;* orcid.org/0000-0001-5015-4225

**Dimitrios Bessas** – *European Synchrotron Radiation Facility, Grenoble F-38043 Grenoble, France;* orcid.org/0000-0003-0240-2540

**Aleksandr Chumakov** − *European Synchrotron Radiation Facility, Grenoble F-38043, France;* orcid.org/0000-0002-0755-0422

**Ilya Sergueev** − *Deutsches Elektronen Synchrotron, Hamburg D-22607, Germany*

**Gabriel E. Büchel** − *ChemConsult GmbH, P.O. Box 43, 9485 Nendeln, Liechtenstein;* orcid.org/0000-0002-5055-7099

**Abdullah A. Al-Kahtani** − *Chemistry Department, College of Science, King Saud University, P.O. Box 2455, Riyadh 11451, Saudi Arabia;* orcid.org/0000-0002-4780-786X

**Liviu Chibotaru** − *Theory of Nanomaterials Group, KU Leuven, Celestijnenlaan 200F, B-3001 Leuven, Belgium;* orcid.org/ 0000-0003-1556-0812

**Notes**
The authors declare no competing financial interest.

**Acknowledgments**

The European Synchrotron Radiation Facility is acknowledged for providing synchrotron radiation beam time at the Nuclear Resonance beamlines ID18 and ID14. The technical assistance of Mr. J.-P. Celse is acknowledged during the beamtime at ESRF. V.B.A. and G.E.B. are thankful to Karl Mayer Stiftung (Triesen, Liechtenstein) and Valüna Stiftung (Vaduz, Liechtenstein) for financial support in purchasing $^{187}$Os metal. L.C. thanks the Researchers Supporting Project no. RSP2024R266, King Saud University, Riyadh, Saudi Arabia. Z.H. and L.U. acknowledge the financial support of research projects A-8000709-00-00, A-8000017-00-00, and A-8001894-00-00 of the National University of Singapore. We are also thankful to Anatolie Dobrov for his help in the synthesis of $^{187}$OsO$_4$ from $^{187}$Os.

**Graphical Abstract**

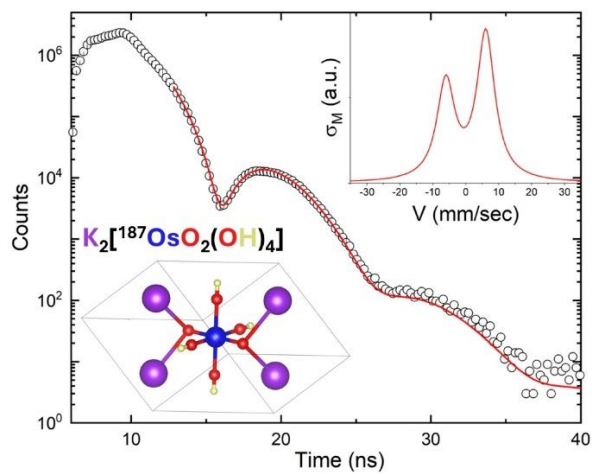